\documentclass{aa}  
\usepackage{txfonts}
\usepackage{epic}
\usepackage{eepic}
\usepackage{graphicx}
\usepackage{epstopdf}
\usepackage{amsmath}
\usepackage{subfig}
\usepackage{xcolor}
\usepackage{todonotes}
\usepackage[]{natbib}
\usepackage{todonotes}
\usepackage{mathtools}

\def\ii{\textrm{i}}
\def\Cref{\overline{C}^{\ \textrm{ref}}}

\begin{document} 
\title{Solar Rossby waves\\ observed in GONG++  ring-diagram  flow maps}
\titlerunning{Solar Rossby waves observed in GONG++} 
   \author{Chris~S.~Hanson \inst{1}, 
          Laurent Gizon\inst{2,3,1}, Zhi-Chao Liang\inst{2}
          }

   \institute{
   \inst{1} Center for Space Science, NYUAD Institute, New York University Abu Dhabi, Abu Dhabi, UAE \\
   \inst{2} Max-Planck-Institut f\"ur Sonnensystemforschung, Justus-von-Liebig-Weg 3, 37077 G\"ottingen, Germany\\
   \inst{3} Institut f\"ur Astrophysik, Georg-August-Universit\"at G\"ottingen, Friedrich-Hund-Platz 1, 37077 G\"ottingen, Germany\\
              \email{hanson@nyu.edu}
             }
             
\date{}

\abstract
  {Solar sectoral Rossby waves have only recently been unambiguously identified in Helioseimsic and Magnetic Imager (HMI) and Michelson Doppler Imager (MDI) maps of flows near the solar surface. So far this has not been done with the Global Oscillation Network Group (GONG) ground-based  observations, which have different noise properties.
 }
   {We utilize 17 years of GONG++ data, to identify and characterize solar Rossby waves using ring-diagram helioseismology. We compare directly with HMI ring-diagram analysis.
   }
{Maps of the radial vorticity are obtained for flows within the top 2~Mm of the surface for 17 years of GONG++. The data is corrected for  {systematic effects including the annual periodicity related to the $B_0$ angle.} 
We then compute the Fourier components of the radial vorticity of the flows in the co-rotating frame. We perform the same analysis on the HMI data that overlap in time.

     }
   {We find that the solar Rossby waves have measurable amplitudes in the GONG++ sectoral power spectra  for azimuthal orders between $m=3$ and $m=15$. 
   The measured mode characteristics (frequencies, lifetimes and amplitudes) from GONG++ are consistent with the HMI measurements in the overlap period from 2010 to 2018 {for $m\le9$. For higher-$m$ modes the amplitudes and frequencies agree within two sigmas.} 
   The signal-to-noise ratio of modes in GONG++ power spectra  {  is comparable to HMI for $8\le m\le11$, but is lower by a factor of two for other modes.}
      }
   {The GONG++ data provide a long and uniform data set to study solar global-scale Rossby waves from 2001.
   }


\maketitle




\section{Introduction}
\citet{loeptien_etal_2018} recently discovered solar global-scale Rossby waves (or r modes) in the SDO/HMI surface flow field using correlation tracking of granulation and helioseismic ring-diagram analysis. The Rossby waves are potentially important as they could be probes of the deep convection zone. \citet{liang_etal_2019} confirmed this result using time-distance helioseismology applied to both HMI and SOHO/MDI observations. It has yet to be seen how well the r modes show in the GONG++ ring-diagram data. This is the goal of this study.


In the first Rossby wave findings, \citet{loeptien_etal_2018} generated radial vorticity maps from flows maps measured by local correlation tracking of granules \citep[LCT,][]{welsch_etal_2004,fisher_welsch_2008} and ring-diagram analysis \citep[RDA,][]{hill_1988}.
They found that the sectoral r-mode spectrum closely follows the standard theoretical dispersion relation $\omega = -2\Omega_{\rm eq}/(m+1)$, where $\Omega_{\rm eq}/2\pi=453.1$~nHz is the equatorial rotation rate and $m$ is the azimuthal order \citep[see, e.g.,][]{saio_1982}. \citet{liang_etal_2019} confirmed the results of \citet{loeptien_etal_2018} using time-distance helioseismology on the meridional component $u_y$ of the horizontal flow near the equator. Unlike \citet{loeptien_etal_2018}, who imaged near surface layers, \citet{liang_etal_2019} imaged at a depth of $0.91$~R$_\odot$. They used 21 years of data spanning both the observation sets of MDI and HMI from 1996 to 2017. Meanwhile, \citet{hanasoge_mandal_2019} provided another independent means of detecting solar r modes, through the use of normal-mode coupling in two years of HMI data. Additionally,  \citet{proxauf_etal_2019} investigated the latitudinal and depth dependence of the r~modes using HMI ring-diagram analysis.
Finally, in an effort to assess the accuracy of machine learning techniques for ring-diagram analysis, \citet{alshehhi_etal_2019} used r modes as a litmus test on the suitability of this new technique for helioseismic inversions. 

In this study we aim to contribute to this growing literature on solar Rossby waves, with an analysis of the 17 years of data from GONG++ and compare a subset of it to the overlapping observations of HMI. We will focus on using the RDA products from both the GONG++ and HMI pipelines. The GONG++ data is interesting for solar r-mode characterization for two reasons. Firstly, the GONG data and HMI data overlap since 2010, enabling the direct comparison of two data sets produced by similar pipelines, but with different noise properties. Secondly, while \citet{liang_etal_2019} merged  the MDI (14 years) and HMI (7 years) data to create a combined 21 years of data, the GONG++ data should be  uniform for 17 years. In this study we explore  the signature  of the solar r modes in the GONG++ data and report differences with the HMI data.

\section{Data analysis and results}\label{sec.dataAnalysis}
\subsection{Ring-diagram analysis}
We use ring-diagram flow maps generated by the GONG++ pipeline from September 2001 to February 2019. The pipeline follows the method outlined in \citet{cobard_etal_2003}, whereby flows are computed from the p-mode frequency shifts extracted from tracked patches (tiles) of the Doppler velocity. These patches are tracked across a transverse cylindrical equidistant projection of the solar disk for approximately 27 hours, following the Snodgrass rotation profile \citep[e.g. Eq.~3 of][]{cobard_etal_2003}. For every day of tracking there are 189 tiles across the solar disk.

For each tile, the two horizontal components of the flows in the $x$ (prograde) and $y$ (northward) directions, $(u_x,u_y)$, are computed as follows:
\begin{itemize}
\item A 2D cosine bell apodization is applied to a $16^\circ\times16^\circ$ square patch, to obtain a circular tile of radius $15^\circ$.
\item For each  data cube of the Doppler velocity $\Psi(x,y,t)$,
a 3D Fourier transform is performed to generate spectral cubes $\widehat{\Psi}(k_x,k_y,\omega)$, from which the power spectrum  $\mathcal{P}(k_x,k_y,\omega) = |\widehat{\Psi}(k_x,k_y,\omega)|^2$ is computed.
\item For each fixed wave number $k=(k_x^2+k_y^2)^{1/2}$, a cylindrical cut through the power spectrum is made  to generate $\mathcal{P}_k(\vartheta, \omega)$, where $\vartheta$ is the angle between the wave vector and the prograde  direction  \citep{bogart_etal_1995}.
\item In the absence of flows, acoustic modes appear as horizontal ridges in $\mathcal{P}_k({\vartheta},\omega)$. The presence of flows causes Doppler frequency shifts that depend on the magnitude and direction of flow. The observed  ridges are fit with a six parameter Lorentzian-like model \citep[e.g.][]{haber_etal_2000}.
\item For each p-mode radial order $n$ and wave number $k$,  two flow parameters $(U_x,U_y)$ are extracted from the  data. The physical flow $(u_x, u_y)$ at a particular depth in the interior is inferred by inverting the set of measured  parameters. In the current study we restrict our attention to the flows at a depth of 2~Mm.
\end{itemize}

\subsection{Sectoral power spectra of radial vorticity}

With the horizontal flow components for each tile computed, we then construct longitude-latitude  maps following the procedure outlined by  \citet{loeptien_etal_2018}. We remove the one year periodicity from each tile which arises primarily from center-to-limb effects that vary with the $B_0$,
{and from a small error in the accepted inclination angle of the solar rotation axis \citep{beck_giles_2005,hathaway_rightmire_2010}}, using Eq.~{A.5} of \citet{liang_etal_2019}.
The model consists of five components, accounting for time-invariant effects and effects with one-year periods.  

With these systematics removed, we then remap the flow maps in a longitude-latitude frame that rotates at the equatorial rotation rate (453.1~nHz), as was done by \citet{loeptien_etal_2018}.   Finally, the radial vorticity, $\zeta$, is computed.

The sectoral power spectrum of the radial vorticity $\zeta$ for the azimuthal order $m$ is computed through
\begin{equation}
    P(m,\omega) = 
    \left|
    \int_{0}^{T}{\rm d}t\int_0^{2\pi}{\rm d}\phi \, 
    e^{-\ii m\phi+\ii\omega t}
    \int_0^{\pi}  {\rm d}\theta \, (\sin\theta)^{m+1}  \zeta(\theta,\phi,t)
    \right|^2 ,
\end{equation}
where $T$ is the observation duration, $\theta$ is the colatitude, and $\phi$ is the longitude. In obtaining the above expression we used the fact that the sectoral spherical harmonic $Y^m_m(\theta,\phi)$ is proportional to  $(\sin\theta)^m e^{\ii m\phi} $.

\begin{figure}[!htb]
    \centering
    \subfloat{\includegraphics[width=0.9\linewidth]{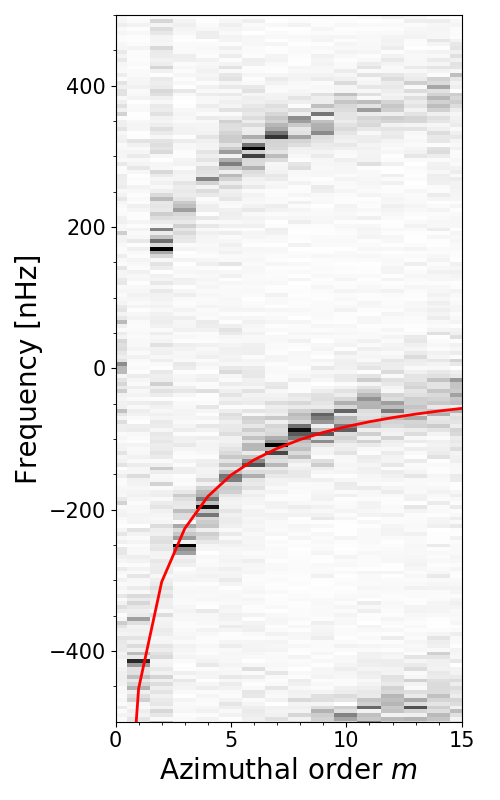}}\\
    \caption{The sectoral power spectrum of the radial vorticity computed from 17.42 years of GONG++ RDA data
    starting 4 September 2001.
   Each $m$ is normalized to the mean power between $-500$ and $100$~nHz. 
    For clarity we plot power spectrum rebinned by a factor 3 in frequency. The red line shows the theoretical dispersion relation for sectoral Rossby waves given in the text. {While a $m=1$ r mode seems to be  present, it is likely low-frequency power leaked from $m=0$.}
    }
    \label{fig.rossby_powspect}
\end{figure}

Figure~\ref{fig.rossby_powspect} shows the power spectrum computed from the 17 years of GONG++ data. We have shown the power spectrum rebinned to one third of the frequency resolution. Similar to previous studies, we clearly identify the r modes which follow the theoretical dispersion relation. 

Interestingly, while $m=2$ mode is absent in the spectrum \citep[see also][for a discussion]{liang_etal_2019}, it appears that there is a peak near the expected frequency of the  $m=1$ mode in Figure~\ref{fig.rossby_powspect}. In order to further investigate this, we perform the same preprocessing on synthetic data (see Appendix~\ref{sec.synthetics}). The synthetics show that any yearly variation signal will leak into $m=1$ due to the partial coverage of the solar surface (window function). Using the annual variation model of \citet{liang_etal_2019}, this leaked signal is removed in our processing. The apparent signal present in $m=1$ of Fig.~\ref{fig.rossby_powspect} coincides with leakage from low frequency power in $m=0$ and is likely not a true $m=1$ r mode. 

\begin{table*}[!htb]
\centering
\begin{tabular}{cccccc}
$m$ & $\omega_0/2\pi$ & $\Gamma/2\pi$ & $S$ & $N$ & $S/N$ \\
 & [nHz] & [nHz] & [$10^{-2}$] & [$10^{-2}$] \\
\hline\hline
3 & $-242.8 \pm 3.1 $ & $30.4^{+7.7}_{-6.2}$ & $60.8^{+23.7}_{-17.0}$ & $3.7^{+0.3}_{-0.3}$ & 16.5\\
4 & $-197.7 \pm 2.7 $ & $24.8^{+6.8}_{-5.3}$ & $71.3^{+30.7}_{-21.5}$ & $3.7^{+0.3}_{-0.2}$ & 19.3\\
5 & $-153.2 \pm 3.7 $ & $29.2^{+9.9}_{-7.4}$ & $33.7^{+15.3}_{-10.5}$ & $4.6^{+0.3}_{-0.3}$ & 7.4\\
6 & $-122.1 \pm 4.4 $ & $51.2^{+11.2}_{-9.2}$ & $50.8^{+15.6}_{-11.9}$ & $4.9^{+0.4}_{-0.4}$ & 10.4\\
7 & $-110.4 \pm 2.7 $ & $28.6^{+6.4}_{-5.3}$ & $164.8^{+62.3}_{-45.2}$ & $5.0^{+0.4}_{-0.4}$ & 32.7\\
8 & $-91.7 \pm 2.7 $ & $30.6^{+6.3}_{-5.2}$ & $171.8^{+61.2}_{-45.1}$ & $3.9^{+0.3}_{-0.3}$ & 44.0\\
9 & $-84.0 \pm 3.2 $ & $35.7^{+7.7}_{-6.3}$ & $98.8^{+33.9}_{-25.2}$ & $4.4^{+0.4}_{-0.3}$ & 22.6\\
10 & $-74.8 \pm 3.0 $ & $28.1^{+7.4}_{-5.9}$ & $106.2^{+43.1}_{-30.7}$ & $6.2^{+0.4}_{-0.3}$ & 17.1\\
11 & $-55.6 \pm 4.0 $ & $36.7^{+10.3}_{-8.1}$ & $49.7^{+19.2}_{-13.8}$ & $5.8^{+0.3}_{-0.3}$ & 8.5\\
12 & $-60.0 \pm 3.2 $ & $26.4^{+8.5}_{-6.5}$ & $48.1^{+21.9}_{-15.1}$ & $4.9^{+0.3}_{-0.3}$ & 9.8\\
13 & $-54.9 \pm 4.8 $ & $46.7^{+12.6}_{-9.9}$ & $22.1^{+7.8}_{-5.7}$ & $3.3^{+0.2}_{-0.2}$ & 6.7\\
14 & $-51.6 \pm 4.9 $ & $38.1^{+13.8}_{-10.1}$ & $13.0^{+5.7}_{-4.0}$ & $2.9^{+0.2}_{-0.2}$ & 4.5\\
15 & $-43.0 \pm 4.5 $ & $44.5^{+11.8}_{-9.3}$ & $14.6^{+5.2}_{-3.8}$ & $1.9^{+0.1}_{-0.1}$ & 7.6\\

\hline
\end{tabular}
\caption{The Rossby mode fit parameters for the GONG++ vorticity power spectrum from 2001 to 2019. {The fit for each mode $m$ was performed in a frequency interval of size 600~nHz, which was stable for our chosen fitting method,
centered on the reference frequency  $-2\Omega_{{\rm eq}}/(m+1)/2\pi$.} The 68\% confidence intervals are included for each fit parameter. {The mode frequencies are measured in the co-rotating frame (rotation rate  $453.1$~nHz).} Negative frequencies indicate retrograde  propagation.
}
\label{tab.mode_details}
\end{table*}

\subsection{Rossby Mode parameters.}
 In this section we seek to characterize the r-mode spectrum. The vorticity power spectrum is fit for each of the Rossby modes $m$ assuming the functional form of the Lorentzian,
\begin{equation}
    \mathcal{L}_m(\omega,\pmb{\lambda}) = \frac{S}{1 + (\omega-\omega_0)^2/(\Gamma/2)^2} + N ,
\end{equation}
where $S$ is the mode amplitude, $\omega_0$ is the mode frequency, $\Gamma$ is the full width at half maximum, $N$ is the background noise of the spectrum and  $\pmb{\lambda}=\{S,\omega_0,\Gamma,N\}$. The probability density function of  the vorticity power spectrum is exponential. As such, the estimates on the parameters $\pmb{\lambda}$ for each mode are determined by minimizing the negative of the  log-likelihood function, 
\begin{equation}
J_m(\pmb{\lambda}) = \sum_{k=1}^K \ln{\mathcal{L}_m(\omega_k,\pmb{\lambda})} + P(m,\omega_k)/\mathcal{L}_m(\omega_k,\pmb{\lambda})  ,
\end{equation}
where $\omega_k$ is  the $k$-th frequency bin within the frequency range of interest, {consisting of $K$ bins}. We use a semi-empirical approach to compute the Hessian and thus the errors on the fit \citep[see][]{toutain_appourchaux_1994}. {The fit for each mode is performed in the frequency range $\pm300$~nHz from the theoretical dispersion relation.}



Table~\ref{tab.mode_details} lists the fit parameters and their error for the 17 year GONG++ RDA r-mode power spectrum. 
The $m=1$ peak is not listed in the table. Its frequency coincides where low frequency $m=0$ power should leak through ($421.41$~nHz) and has a much  smaller amplitude than the other modes. The fit for  $m=2$ was not performed due to the absence of any signal.

\subsection{Comparing GONG++ and HMI}

\begin{figure*}[!htb]
    \centering
    \includegraphics[width=0.95\linewidth]{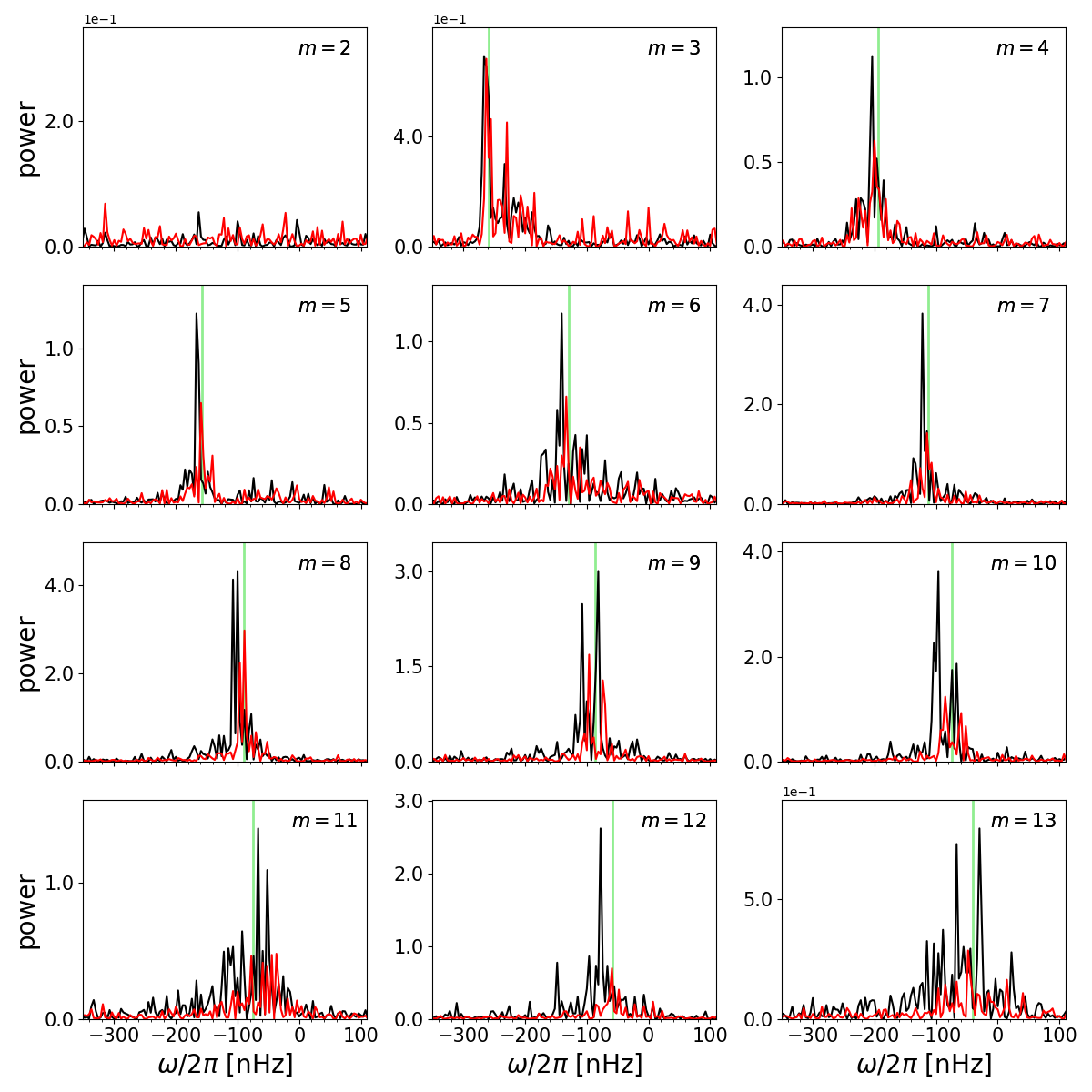}
    \caption{Comparisons of the GONG++ (red) and HMI (black) Rossby wave power spectra for the same observation period of 8 years {(2010-2018)}, using maps of  radial vorticity inferred from ring-diagram analysis. The green vertical lines show the measured mode frequencies of \citet{loeptien_etal_2018} from granulation tracking data.
    }
    \label{fig.comparing_gonghmi_spectra}
\end{figure*}

In this section we compare the r-mode power spectrum derived from the respective RDA flow map pipelines of HMI and GONG++, for a shared observation period between May 2010 and December 2018. Figure~\ref{fig.comparing_gonghmi_spectra} compares the GONG++ and HMI r-mode power spectra for the shared observational period. These results show that while the noise is different, the mode power spectra are in general agreement. The amplitudes of GONG++ and HMI r-mode signals are close, with the GONG++ having a greater background noise. For $m>8$ the mode power in the GONG++ data becomes noticably smaller. 


\begin{figure*}[!htb]
    \centering
    \subfloat{\includegraphics[width=0.5\linewidth]{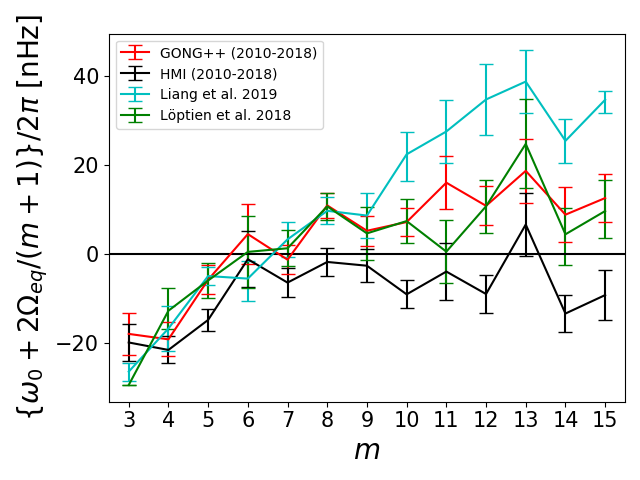}}
    \subfloat{\includegraphics[width=0.5\linewidth]{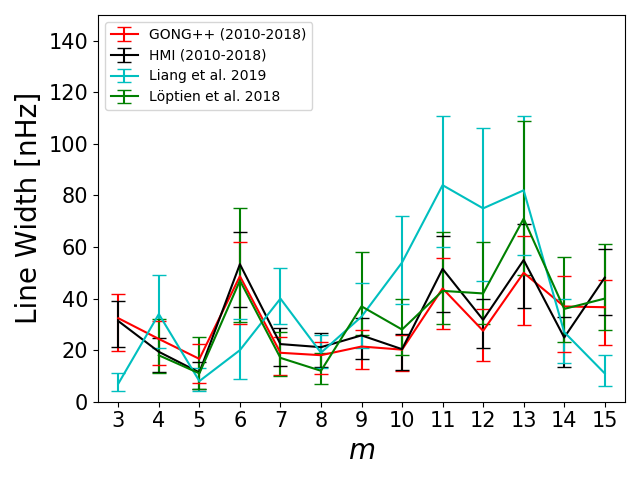}}\\
    \subfloat{\includegraphics[width=0.5\linewidth]{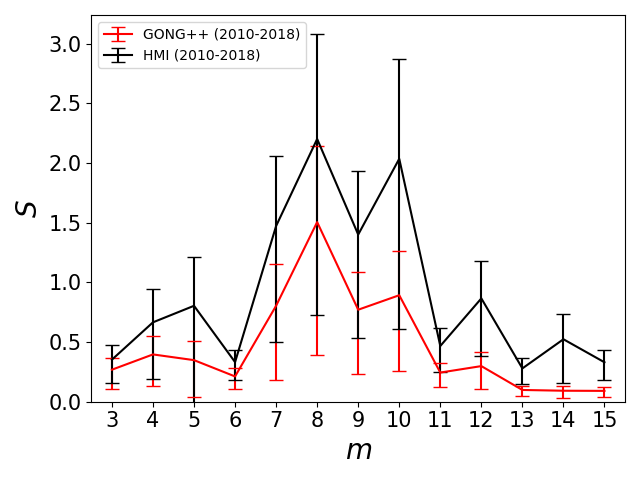}}
    \subfloat{\includegraphics[width=0.5\linewidth]{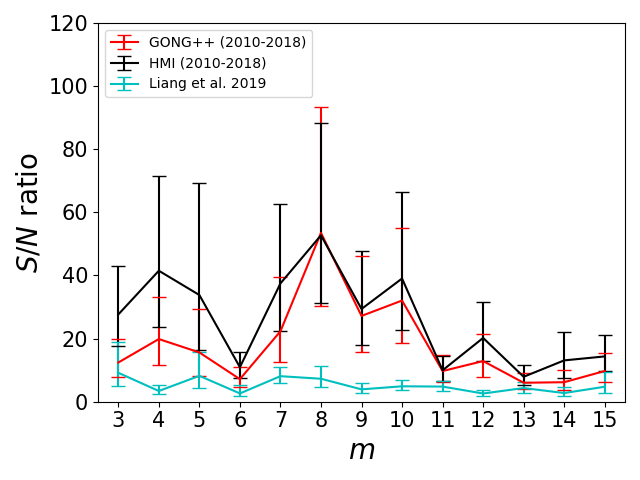}}
    \caption{
    Top left: Difference between measured and theoretical resonant frequencies for different data sets:
 GONG++ RDA (red, {2010-2018}), HMI RDA (black, {2010-2018}), MDI+HMI time-distance from \citet{liang_etal_2019} (cyan, {1996-2017}) and HMI LCT from \citet{loeptien_etal_2018} (green, {2010-2016}).
    Top right: Measured line widths. Bottom left: Measured vorticity amplitudes. Bottom right: $S/N$ ratios for GONG++ and HMI ring diagrams.  Error bars are one sigma.
}
    \label{fig.comparing_gonghmi}
\end{figure*}

Figure~\ref{fig.comparing_gonghmi} compares the measured frequencies, line widths, vorticity amplitudes and $S/N$ ratios between four different data sets. The data sets include GONG++ and HMI ring diagrams with overlapping observation time, the travel-time results of \citet{liang_etal_2019} derived from time-distance helioseismology, and the LCT results of \citet{loeptien_etal_2018}. These results show that within error bars the reported mode frequencies and line widths agree for $m\le9$. For $m\ge10$ the mode frequencies tend not to agree, with the time-distance results of \citet{liang_etal_2019} being shifted towards smaller negative frequencies, while the HMI RDA data analyzed here tends to have greater negative frequencies. 
In terms of line width, the four spectra tend to be in agreement for all $m$, though the large reported errors make it difficult to make any strong conclusions on this characteristic of the modes. 
{The HMI and GONG++ RDA mode amplitudes are within error limits for $m\le10$. But, for higher order modes HMI data has a greater amplitude by a factor two. }
{Comparing the $S/N$ ratios, shows that in general the ratio is greater in HMI RDA than GONG++ RDA by a factor of two for $m<8$, are similar from $m=8$ to $m=11$, and are greater again for higher order modes. Our results also show that the $S/N$ ratios  are significantly higher in ring-diagram data than in travel-time measurements \citep{liang_etal_2019}.}

\section{Discussion and conclusions}
We have utilized 17 years of observations from the ground based GONG program to characterize solar equatorial Rossby waves. Using ring diagram analysis from the GONG++ pipeline, we have clearly identified the solar Rossby waves in the sectoral power spectra of the  radial vorticity maps. 

Like previous studies we have found that the $m\le2$ r modes are absent from the data. An apparent $m=1$ peak appears in the power  spectra but we conclude that this peak is leaked through from low frequency power in $m=0$. The synthetics show that a dipole r mode will appear as two peaks separated by $2\times 31.7$~nHz due to leakage. No such configuration is seen in the solar spectrum. 



With 17 years of near continuous data, we measured the Rossby mode characteristics. 
We find that our results, for the period during which GONG++ and HMI data overlap, agree with both \citet{loeptien_etal_2018} and \citet{liang_etal_2019} for $4\le m\le9$. For $m>9$ modes the GONG++ r-mode frequencies are only in agreement with \citet{loeptien_etal_2018}. {This is due to the chosen method for computing the r-mode spectrum. We have found that if we also compute the spectrum through the Fourier transform of $u_y$ at the equator, the frequencies agree with \citet{liang_etal_2019}. These discrepancies arise for high $m$, because the latitudinal eigenfunctions of the r-modes are not sectoral spherical harmonic functions \citep{proxauf_etal_2019}. Our analysis also suggests that this choice in latitudinal basis function contributes to the small discrepancy for $m=3$. }
The amplitudes from the GONG++ and HMI RDA are within errorbars for $m\le10$. For higher-order modes the HMI RDA amplitudes are greater than GONG++ by a factor of two. 
The measured $S/N$ ratios of modes in GONG++ RDA power spectra are in the range between 5 and 45. This shows the suitability of the GONG++ data for future studies of global-scale  Rossby waves.
We note we do not detect any mode in the $m=2$ sectoral power spectrum, which is coherent with previous studies.

In this study we have focused on only one depth (2~Mm), and measured the mode characteristics for the entire 17 (or 8) year time series. We have found good agreement with HMI, despite the lower $S/N$ ratios in the GONG++ data. The advantage that GONG++ data has over other data sets is the long and uniform observation window. The next stages for Rossby wave studies should focus on the temporal variation of the waves. With the long time series of GONG shown in this study, and the combined HMI and MDI time series of \citet{liang_etal_2019}, the temporal dependence of r-modes could be investigated with two independent data sets.

\appendix

\section{Synthetics}\label{sec.synthetics}
In order to quantify the effects of systematics, we generate synthetic Rossby waves and analyze the effects of our data processing (Sec.~\ref{sec.dataAnalysis}) and the window function due to partial coverage of the Sun. For the functional form of the Rossby waves, we assume they are purely horizontal and obey mass conservation. The flow components $(u^m_\theta,u^m_\phi)$ of a sectoral r mode of azimuthal order $m$ are defined by,

\begin{align}\label{eq.rossbySyn}
    \begin{split}
   u_\theta^m(\theta,\phi,t)&= -A (\sin\theta)^{m-1} \sin(m\phi-\sigma_m t) e^{-\Gamma t},\\
         u_\phi^m(\theta,\phi,t) &= -A\cos\theta\   (\sin\theta)^{m-1} \cos( m \phi -\sigma_mt) e^{-\Gamma t},\\
    \end{split}
\end{align}
where $\sigma_m=-2\Omega_{\rm eq}/(m+1)$ and $A=1$~m/s. Here, $\Gamma/2\pi$ is chosen to be 10~nHz.
{For a single ring-diagram tile the flow components $u_x$ and $u_y$ can be identified with $u_\phi$ and $-u_\theta$, respectively. The horizontal flow field within a single tile centered at co-latitude $\theta$ and longitude $\phi$ is then computed by summing the individual r modes:}
\begin{align}
    u_x(\theta,\phi,t) &= \sum_{m=1}^{M} {u_\phi^m(\theta,\phi,t)}, \\
    u_y(\theta,\phi,t) &= - \sum_{m=1}^{M} {u_\theta^m(\theta,\phi,t)}. 
\end{align}
The radial vorticity $\zeta$ is then computed through;
\begin{equation}
    \zeta(\theta,\phi,t) =   \frac{1}{R_\odot\sin\theta}\left[ \frac{\partial}{\partial \theta}(u_x\sin\theta) + \frac{\partial}{\partial \phi}u_y\right]
    + \frac{A}{R_\odot}\cos({\omega_\oplus t}) ,
\end{equation}
where we added a signal with a yearly period ($2\pi/\omega_\oplus=1$~year) to emulate the annual systematic error that is present in RDA \citep[e.g.][]{komm_etal_2015}. 
We limit ourselves to a sum of $M=15$ modes.

Using Eq.~\ref{eq.rossbySyn}, we compute the flows in each $15^\circ$ tile, across the solar surface. We then examine three cases: 1) An ideal case where we have tiles across the entire solar surface at all times. 2) The realistic case when we only have tiles observed by GONG on the visible disk, but no annual variation is removed. 3) The same as the previous case, except after the performing the processing outlined in this paper (e.g. temporal systematic removal).

\begin{figure}[!htb]
    \centering
    \includegraphics[width=\linewidth]{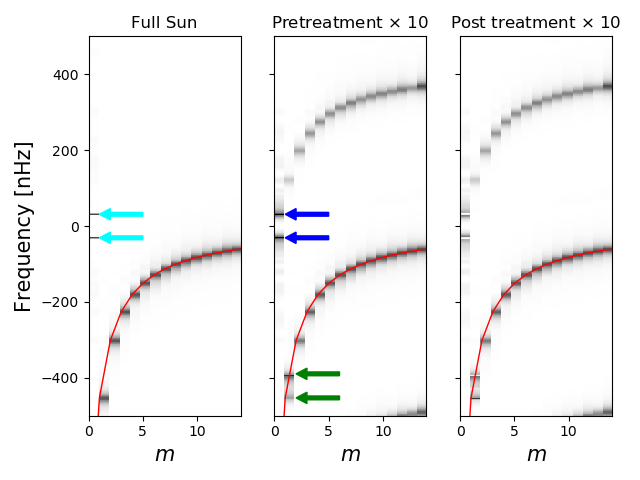}
    \caption{{Sectoral} power spectra for the synthetic {radial vorticity data} for the complete solar surface (full Sun, left panel), including the observational window function (pre-treatment, central panel), and after processing (post-treatment, right panel).  Cyan arrows show the peaks at $\pm (1\; {\rm year})^{-1}=\pm 31.7$~nHz. Blue arrows show the leakage of $|m|=1$ into $m=0$ once the window function is applied. These leakages coincide with the annual variation. After treatment, the annual variation and its leakage disappears, leaving the $m=1$ r-modes and associated leakage.  The green arrows highlight the effects the window has on $m=1$ power. }
    \label{fig.windowAnalysis}
\end{figure}

Figure~\ref{fig.windowAnalysis} shows the {sectoral} power spectra {of the radial vorticity} for these three cases. In the ideal case, where the entire sun is observed, we see a clean spectrum without aliasing. 
In the realistic case (2), there is leakage of modes into their neighbouring modes, albeit at frequencies $\pm 421.54$~nHz from the central frequency. For the ridge of concern to us, the most significant effect is the leakage into $m=1$.


\begin{acknowledgements}
This work was supported by NYUAD Institute Grant G1502. LG acknowledges partial support from the European Research Council Synergy Grant WHOLE SUN 810218. This work utilizes data obtained by the Global Oscillation Network Group (GONG) Program, managed by the National Solar Observatory, which is operated by AURA, Inc. under a cooperative agreement with the National Science Foundation.  The HMI data is courtesy of NASA/SDO and the HMI Science Team. 
CSH thanks Frank Hill and Andrew Marble for their assistance with the GONG data. We also  thank  Bastian Proxauf and Jishnu Bhattacharya for insightful discussions and Martin Bo Nielsen for performing consistency checks between the analytical fitting of this study and Markov Chain Monte Carlo. 
\end{acknowledgements}

\bibliographystyle{aa}
\bibliography{References}

\end{document}